# Seamless monolithic three-dimensional integration of single-crystalline films by growth

Ki Seok Kim[1,4†], Seunghwan Seo[1,4†], Junyoung Kwon[2,3†], Doyoon Lee[4,8†], Changhyun Kim[2,3†], Jung-El Ryu[1,4], Jekyung Kim[1,4], Min-Kyu Song[1,4], Jun Min Suh[1,4], Hang-Gyo Jung[5], Youhwan Jo[6], Hogeun Ahn[5], Sangho Lee[1,4], Kyeongjae Cho[6], Jongwook Jeon[5], Minsu Seol[2,3]*, Jin-Hong Park[5,7]*, Sang Won Kim[2,3]*, and Jeehwan Kim[1,2,3,4,8]*

[1]Department of Mechanical Engineering, Massachusetts Institute of Technology, Cambridge, MA 02139, USA

[2]2D Device Laboratory, Samsung Advanced Institute of Technology, Suwon, Korea

[3]Device Research Center, Samsung Advanced Institute of Technology, Suwon, Korea

[4]Research Laboratory of Electronics, Massachusetts Institute of Technology, Cambridge, MA 02139, USA

[5]Department of Electrical and Computer Engineering Sungkyunkwan University, Suwon-si, South Korea

[6]Department of Materials Science and Engineering, The University of Texas at Dallas, Richardson, TX 75080, USA

[7]SKKU Advanced Institute of Nano Technology (SAINT), Sungkyunkwan University, Suwon-si, South Korea.

[8]Department of Materials Science and Engineering, Massachusetts Institute of Technology, Cambridge, MA 02139, USA

[†]Equally contributed / *Correspondence to minsu.seol@samsung.com, jhpark9@skku.edu, swon80.kim@samsung.com, jeehwan@mit.edu

The demand for the three-dimensional (3D) integration of electronic components is on a steady rise. The through-silicon-via (TSV) technique emerges as the only viable method for integrating single-crystalline device components in a 3D format, despite encountering significant processing challenges. While monolithic 3D (M3D) integration schemes show promise, the seamless connection of single-crystalline semiconductors without intervening wafers has yet to be demonstrated. This challenge arises from the inherent difficulty of growing single crystals on amorphous or polycrystalline surfaces post the back-end-of-the-line process at low temperatures to preserve the underlying circuitry. Consequently, a practical growth-based solution for M3D of single crystals remains elusive. Here, we present a method for growing single-crystalline channel materials, specifically composed of transition metal dichalcogenides, on amorphous and polycrystalline surfaces at temperatures lower than 400 °C. Building on this developed technique, we demonstrate the seamless monolithic integration of vertical single-crystalline logic transistor arrays. This accomplishment leads to the development of unprecedented vertical CMOS arrays, thereby constructing vertical inverters. Ultimately, this achievement sets the stage to pave the way for M3D integration of various electronic and optoelectronic hardware in the form of single crystals.

The integration of three-dimensional (3D) electronics has become a critical aspect of modern electronic industries due to the limitations of scaling current nanoscale devices[1–3]. Additionally, arranging chips vertically can significantly reduce resistive-capacitive (RC) delays in integrated circuitry, leading to lower power consumption and more efficient data exchange within system-on-chip (SoC) designs[4]. Moreover, system-on-chip components can be more flexibly accommodated within a smaller footprint. The most concise and effective way to establish connections between electronic devices can be achieved through monolithic 3D (M3D) integration[5]. In this approach, the uppermost "single-crystalline" devices are interconnected without the need for thick wafers. However, it is important to note that in order to attain exceptionally high-performing single-crystalline devices, the use of single-crystalline wafers is essential. Thus, conventional 3D integration techniques typically keep the silicon wafer and employ a through-silicon-via (TSV) approach, where micron-scale holes are drilled entirely through the wafer, followed by bonding TSV-processed wafers[6,7]. However, there are critical challenges associated with this TSV technology such as costly hole drilling processes, chip misalignment, and trading valuable chip spaces with TSVs. However, so far, TSVs have been an only viable way to connect single-crystalline devices as direct epitaxial growth of single-crystalline devices on amorphous back-end-of-line (BEOL) layers is impossible. As a result, the significant potential of monolithic integration of single-crystalline devices through direct growth has not yet been proven, despite its substantial potential impact.

As an alternative approach, it is possible to detach single-crystalline channel from the wafer and transfer them onto finished chip dies to achieve wafer-free M3D integration. One notable demonstration of this concept is the CoolCube™ technique developed by LETI, where a silicon-on-insulator structure is transferred onto BEOL finished wafers, followed by the full integration of logic circuitry[8,9]. However, it's important to note that this method still requires a wafer bonding step. Furthermore, there is a significant challenge associated with the activation of source/drain regions, which typically requires temperatures above 600 °C. This high-temperature process can severely degrade the underlying circuitry, making it essential to maintain a process temperature below 400 °C for preserving the integrity of the integrated components[10]. Another potential approach involves the room temperature transfer of devices that have been fully integrated at a high temperature on the donor wafer[11,12]. However, this method raises the issue of aligning nanoscale devices precisely onto the underlying circuitry, which can be a complex and precise task. Ultimately, the ideal solution would involve the direct growth of single-crystalline channel materials on amorphous BEOL layers at a temperature below 400 °C followed by device integration. Nevertheless, it is recognized that accomplishing such a task is nearly unfeasible.

Here, we present our first-ever demonstration of single-crystalline growth of channel materials at 385 °C on an amorphous oxide layer-coated silicon wafer. This breakthrough enables the seamless M3D integration of single-crystalline channel materials. An n-type metal–oxide–semiconductor (MOS) arrays have been successfully integrated by the growth on top of amorphous layers encapsulating single-crystalline p-type MOS arrays. As a result, we have successfully created an unprecedented single-crystalline 3D stacked field-effect transistors (3DS FET) arrays, which has the potential to double the integration density of logic circuits, representing a significant advancement in semiconductor technology. 3DS FETs are also known as complementary FETs (CFETs) and in this paper we will call it as vertical complementary metal–oxide–semiconductor (CMOS), hereafter. We have adopted a two-dimensional (2D) transition metal dichalcogenides (TMDs) as a channel material because of the following reasons: i) it is considered a highly promising alternative to silicon for advanced node transistors, primarily because it effectively mitigates the short-channel effects[13,14], ii) all post-growth fabrication processes occurring at the temperatures lower than 400 ºC justify its strong potential for M3D integration, and iii) geometric confinement during growth has the potential to facilitate the formation of single-crystalline TMDs on amorphous surfaces[13]. Our research employs confined selective growth, enabling the formation of single-crystalline TMDs within confined areas by promoting a single nucleation event. We make efficient use of the edges and corners of these confined trenches as heterogeneous nucleation sites, thus allowing to successfully grow perfect single-crystalline $MoS_2$ and $WSe_2$ on amorphous insulation layers at sub-400 °C. This makes a clear contrast to our previous demonstration at very high growth temperatures, typically ranging from 700 to 900 °C[13]. This achievement allows us to integrate single-crystalline nMOS directly on top of completed pMOS arrays by the growth. The underlying devices maintain their performance integrity after top MOSFET fabrication, allowing seamless M3D integration. As a result, we have successfully fabricated vertical CMOS and constructed inverters that exhibit excellent performance compared to that of TMD-based CMOS previously demonstrated by mechanical stacking, even at a relatively long channel length of 400 nm. Our simulations further suggest that once advanced node CMOS technology is developed using our TMD-based vertical CMOS, the performance will surpass that achieved with silicon if the interfaces are maintained pristine. This work establishes a foundation for advancing Moore's law by doubling the integration density of logic transistors. Additionally, the successful demonstration of growing single-crystalline devices on top of finished circuitry holds the promise of true wafer-free vertical M3D integration of electronics and photonics in the future.

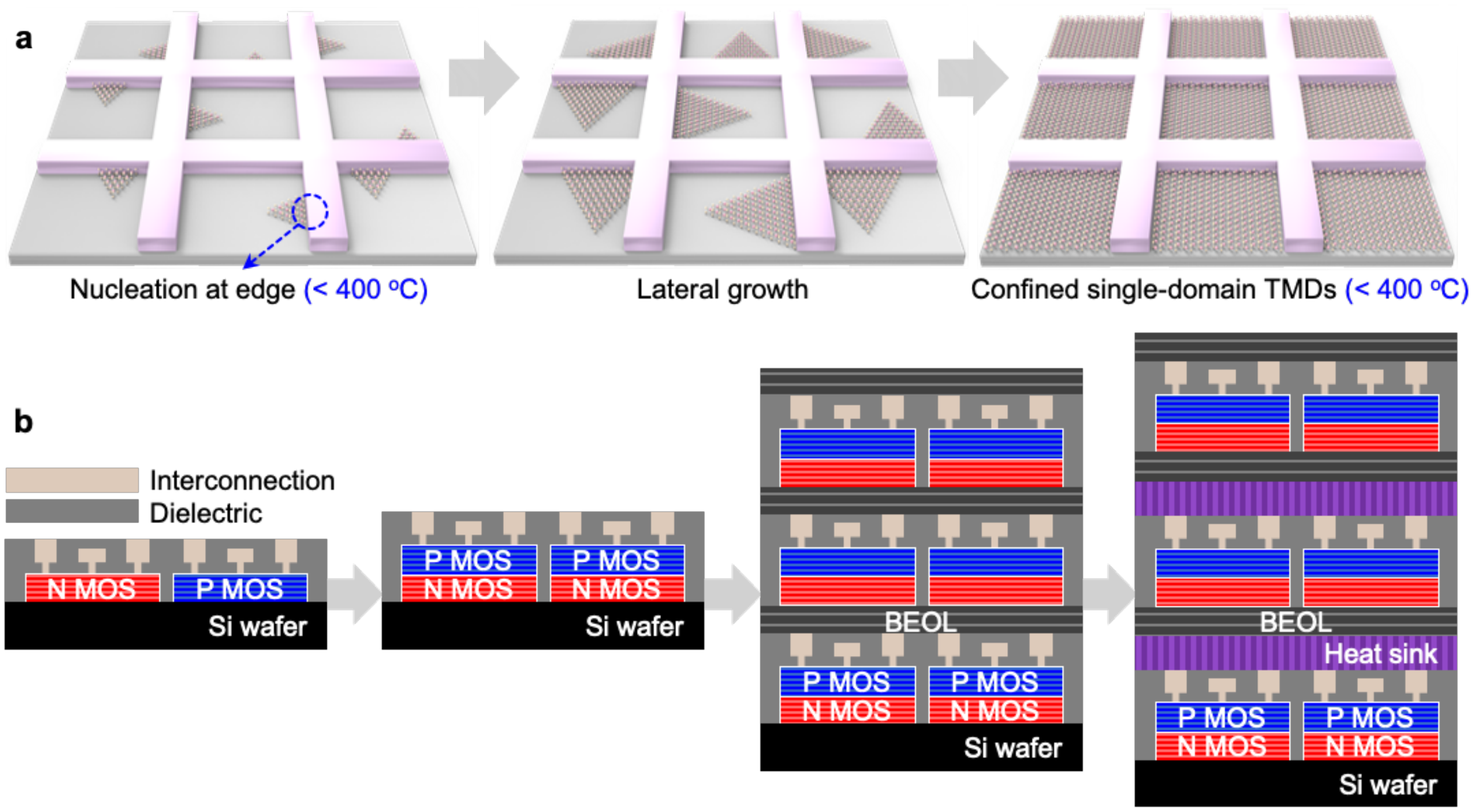

**Fig. 1 | Low-temperature growth of single-crystalline film for seamless M3D integration. a,** Schematics illustrating low-temperature (<400 °C) growth of single-crystalline TMD array, highlighting tendency of nuclei to form at edges or corners of patterned structure. **b,** Schematic diagrams showing planarly (1st schematic) and vertically (2nd schematic) integrated CMOS elements and its M3D integration based on BEOL (3rd schematic) and with heak sink (4th schematic).

Significant efforts have been dedicated to reducing the growth temperature of TMDs to preserve the performance of the underlying electronic device circuitry when adapting TMD channel to M3D integration. Achieving this has involved techniques like breaking down growth precursors at higher temperatures while maintaining a cooler growth zone or using surfactants to extend the diffusion length of adatoms[15–18]. However, TMDs grown at lower temperatures tend to result in small grains in their polycrystalline films with electronic properties that are far less than ideal. Thus, the key is to obtain TMD films with a larger grain size, ideally a single domain at a low temperature while it has not been demonstrated. Interestingly, the research community has not given as much attention to strategies for promoting nucleation, even though nucleation is the most crucial factor in initiating the growth process. According to classical nucleation theory, when the growth temperature is high enough to exceed the activation energy necessary for homogeneous nucleation, nucleation can take place uniformly across the surface. However, at lower temperatures, which do not provide sufficient energy to overcome the activation barrier for homogeneous nucleation, nucleation events are constrained by kinetics. This results in a preference for heterogeneous nucleation at the edges or corners[19]. We have applied the same principle to reduce the nucleation temperature by promoting heterogeneous nucleation at the confined growth geometry as shown in Fig. 1a. We have utilized $SiO_2$ selective growth masks for TMDs on amorphous (a)-$HfO_2$ surfaces and guided nucleation at the edges and corners of $SiO_2$ masks as sites for heterogeneous nucleation. Consequently, even at low temperatures that would not typically allow nucleation on the flat surface, nucleation can still take place, enabling the deposition of TMD thin films. Simultaneously, we have carefully designed the distribution and size of selective growth trenches to ensure the formation of single nuclei at a single trench. The trench size is kept small enough to complete the lateral growth of TMDs before a second nucleation event occurs, resulting in the formation of single-domain TMDs on a-$HfO_2$ coated Si wafer. Thus, finally wafer-scale single-crystalline TMDs on dielectric layers could be grown on the Si wafer at temperatures below 400 °C, a critical temperature at which modern electronic circuitry based on Cu interconnects can operate reliably. This allows to demonstrate unprecedented single-crystalline circuitries that are vertically integrated by the direct growth as illustrated in Fig. 1b. This outcome lays strong foundation to vertically stack CMOS structures furthermore to maximize the density of logic transistors within a given space or seamless monolithic integration of logics with memories. Simultaneously, heat spreading layers could also be inserted adjacent to the circuitry that potentially generates heat. Thus, this allows the continuation of Moore's Law and the vertical integration of high-bandwidth memories while the new cooling scheme suitable for this M3D must be developed in the future.[20,21]

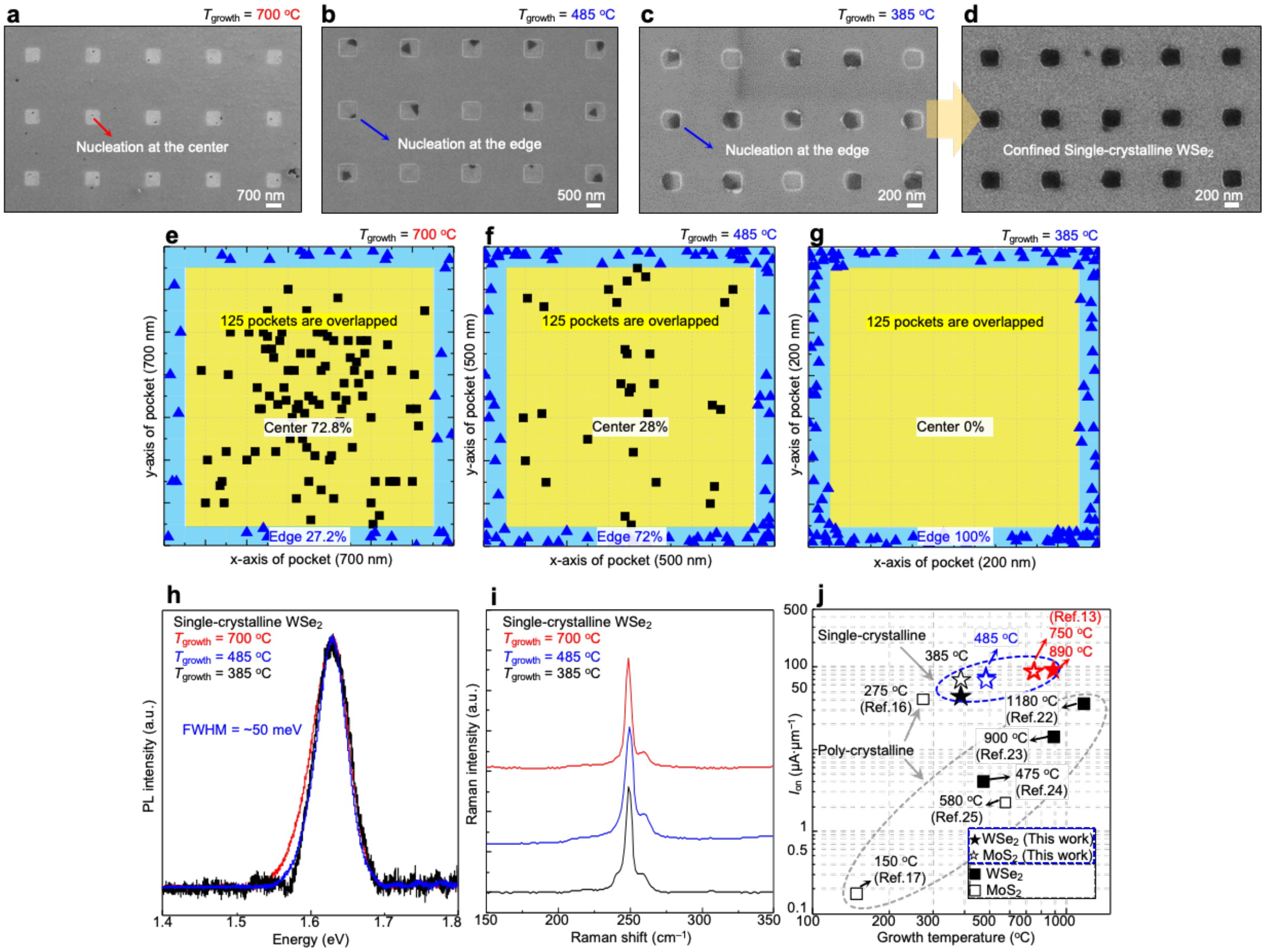

**Fig. 2 | Low-temperatuer growth of single-crystalline $WSe_2$. a–c,** SEM images depicting initial nucleation of single-crystalline $WSe_2$ within patterned $SiO_2$ pockets formed on $HfO_2$ substrate under $T_{growth}$ of 700 °C (**a**), 485 °C (**b**), and 385 °C (**c**) conditions, respectively, highlighting increase in initial nucleation probability near edge of pocket with decreasing $T_{growth}$. **d,** SEM image of single-crystalline confined $WSe_2$ grown at 385 °C. **e–g,** Statistical analysis of initial nucleation formation locations observed in 125 pockets, verifying nucleation formation with probabilities of 27.2% at 700 °C (**e**), 72% at 485 °C (**f**), and 100% at 385 °C (**g**) near the edge, respectively. **h–i,** PL analysis (**h**) and Raman analysis (**i**) on single-crystalline $WSe_2$ grown at 700 °C (red-colored profile), 485 °C (blue-colored profile), and 385 °C (black-colored profile). **j,** $I_{on}/W_{ch}$ of CVD-grown $WSe_2$ and $MoS_2$ channel-based pMOS and nMOS transistors with respect to $T_{growth}$ of $WSe_2$ and $MoS_2$ channel layer.[13,16,17,22–25]

In order to verify the nucleation dynamics on structured surfaces, we conducted density functional theory (DFT) calculations and our DFT analysis revealed that the amorphous nature of $HfO_2$ at temperatures below 400 °C enhances nucleation at the edges of the trenches. The calculation illustrates that the binding of TMDs on amorphous a-$HfO_2$ is notably weaker than on crystalline (c)-$HfO_2$ (Extended Data Fig. 1a). Consequently, nucleation at the edges of $SiO_2$ is further stimulated, resulting in a 35% increase in binding energy at the edges (Extended Data Fig. 1b). Fig. 2a shows experimental results of the nucleation tendencies of $WSe_2$ at 700 °C. Statistical analysis reveals that 72.8% of nuclei are formed directly on the planar $HfO_2$ surface at this temperature (see Fig. 2e). It should be noted that every trench has a single nucleus and 125 trenches are overlapped for the statistics. Intriguingly, dramatic transition of nucleation from the center to the edge occurs at 485 °C (Fig. 2b,f). The statistics were randomly selected from a 1.5 by 1.5 cm area (Extended Data Fig. 2). Finally, all nuclei form at the edges of $SiO_2$ trenches at 385 ºC (Fig. 2c,g). In contrast, non-patterned areas do not exhibit nucleation and consequently do not form any films (Extended Data Fig. 3). As shown in Fig. 2d, further growth at such low temperatures results in the production of confined single-domain TMDs, as the trench size remains small enough to complete lateral TMD growth within a very short timeframe before secondary nucleation occurs. Notably, the quality of TMDs remains even at 385 °C due to their single-crystalline nature, and this temperature is low enough to preserve the performance of modern electronic logic and memory circuitry. As shown in Fig. 2h,i, the photoluminescence (PL) and Raman spectra remain stable up to 385 °C. As shown in Fig. 2j,[13,16,17,22–25] the on-current per channel width ($I_{on}/W_{ch}$) of FETs fabricated using our single-crystalline TMDs is preserved up to 485 °C, with a slight degradation observed from around 385 °C (approximately 25%). This makes a contrast to the previous demonstrations to grow polycrystalline TMDs at low temperatures, in which the mobility degradation was unavoidable due to reduced grain sizes. Since the $HfO_2$-coated Si substrate did not exhibit leakage current caused by crystallization below 500 ºC, we employed $WSe_2$ grown at 485 ºC, which has relatively good electrical performance, as pMOS. (see Extended Data Fig. 4 and Extended Data Fig. 5).

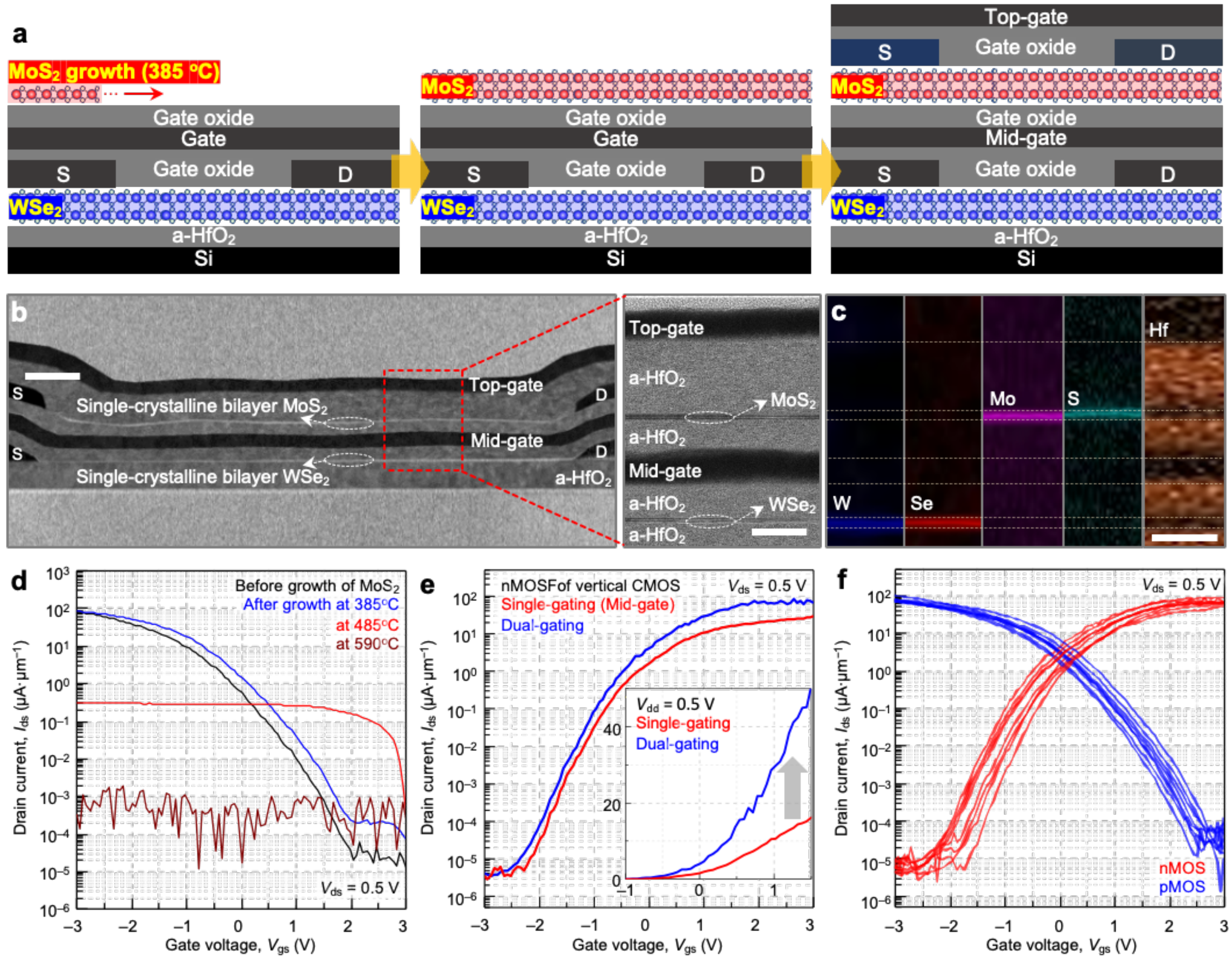

**Fig. 3 | Seamless M3D integration. a,** Schematic diagrams showing seamless M3D integration process of single-crystalline vertical CMOS device, which includes following core steps. First step is low-temperature growth of single-crystalline $MoS_2$ layer on top of single crystalline pMOS. Following step is integration of single crystalline nMOS, demonstrating seamless M3D integration. **b,** Cross-sectional scanning transmission electron microscopy image showing the structure of vertical CMOS. **c,** EDS mapping results showing signals related to W, Se, Mo, S and Hf elements. **d,** Transfer characteristics of pMOS before (black-colored) and after growth of single-crystalline $MoS_2$ layer, where the growth temperature is 385 °C (blue-colored), 485°C (red-colored), and 590 °C (wine-colored), respectively. **e,** Transfer characteristics of nMOS when single- (red-colored) or dual-gating (blue-colored) is applied. **f,** Transfer characteristics of vertical CMOS array, where red-colored and blue-colored curves denote that of nMOS and pMOS, respectively.

To showcase seamless M3D integration of single-crystalline devices enabled by low temperature non-epitaxial growth of single-crystalline semiconductors, we have constructed unprecedented vertical CMOS, also known as “3DS FET” or “CFET”, with the potential to reduce the footprint of logic integration by half, effectively doubling logic integration density. First, single-crystalline pMOS arrays were fabricated by growing single-crystalline $WSe_2$ on an a-$HfO_2$ coated Si substrates followed by finishing S/D contacts and gate stacks. The first pMOS arrays were isolated by an a-$HfO_2$ encapsulation. Then, a vertical CMOS was finished by constructing nMOS based on single-crystalline $MoS_2$ by growing it directly on a-$HfO_2$ encapsulation layer at 385 °C. (Fig. 3a and Extended Data Fig. 6). Actual images of this single-crystalline vertical CMOS were displayed in Fig. 3b which was taken by a cross-sectional high-resolution transmission electron microscopy (HRTEM). The energy-dispersive spectrometer (EDS) examination confirms the atomic composition of each layers in a such vertical CMOS (Fig. 3c). We have conducted a sequential analysis of the electrical properties of both the lower pMOS and upper nMOS. This analysis involved i) an evaluation of the influence of $MoS_2$ growth temperature on the underlying $WSe_2$ pMOS and ii) a comprehensive performance assessment of the top nMOS. Fig. 3d demonstrates a transfer characteristic of underlying $WSe_2$ pMOS after the growth of the single-crystalline $MoS_2$. As seen in the plot, the pMOS before the $MoS_2$ growth exhibits on-current ($I_{on}/W_{ch}$) of 82.9 μA/μm at the channel length of 400nm and $V_{ds}$ = 0.5 V, while obtaining high on/off current ratio reaching as high as $6.59\times10^6$. After the growth of the single-crystalline $MoS_2$ layer at 385 °C, the transfer characteristics of the $WSe_2$ pMOS remains unaffected. However, the performance of underlying pMOS has been severely degraded when nMOS channel is grown at 485 and 590 °C (see Fig. 3d and Extended Data Fig. 7). Thus, we proceeded to construct a vertical CMOS with $MoS_2$ growth at 385 °C. However, the on-current of the nMOS is approximately 56% less than that of the pMOS. Thus, to match the current, we have applied dual gate bias on $MoS_2$ nMOS which enhances the $I_{on}$ performance of nMOS, thereby reducing the current mismatch to less than 10% as shown in Fig. 3e. The transfer curves of matching nMOS and pMOS arrays have been illustrated in Fig. 3f and statistics shows an average variation of the on-current around 15%.

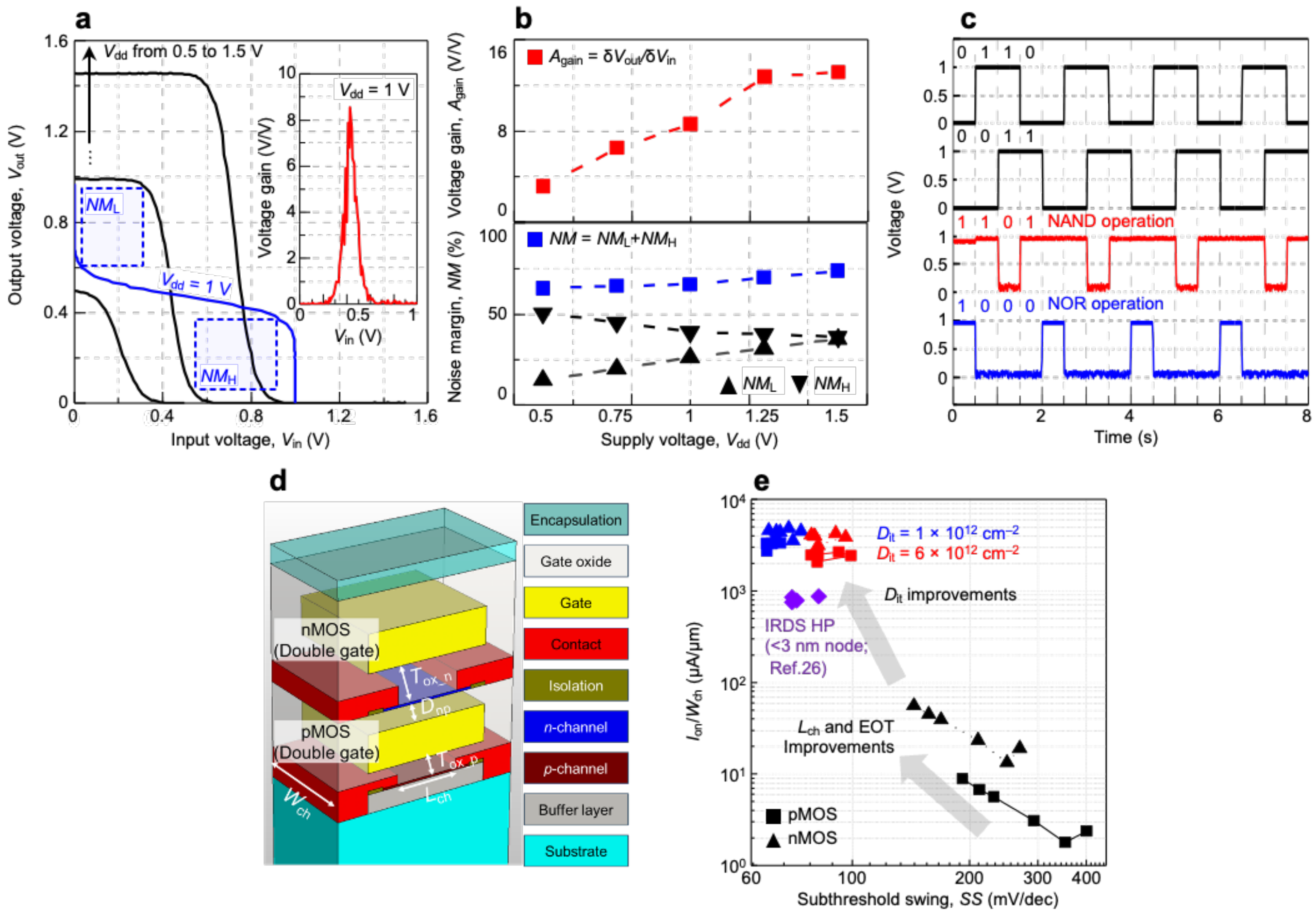

**Fig. 4 | M3D integrated CMOS-based logic circuits and advanced CMOS process node application. A,** Voltage transfer characteristics of vertical CMOS inverter, where inset graph and blue-colored boxes denote voltage gain profile and noise margin ($NM_L$ and $NM_H$), respectively. **b,** Extracted voltage gain (top panel) and estimated noise margin (bottom panel) values with respect to supply voltage ranging from 0.5 to 1.5 V, which are extracted from voltage transfer characteristics. **c,** Timing diagram of vertical CMOS-based NAND (red-colored line) and NOR (blue-colored line) gates. **d,** Structural schematic of M3D integrated CMOS used in 3D FEM simulation. **e,** $I_{on}/W_{ch}$ with respect to $SS$ of fabricated vertical CMOS (black-colored plots) and vertical CMOS when $D_{it}$ is assumed to be less than $6 \times 10^{12}$ cm$^{-2}$ (red- and blue-colored plots). Targeted high-performance metrics including $I_{on}/W_{ch}$ and $SS$ at 3 nm, 2 nm, 1.5 nm, and 1 nm process node that is provided in IRDS 2022[26] (purple-colored plots).

With an established vertical CMOS, we have constructed "monolithic" vertical inverters, and their evaluation is depicted in (Fig. 4a and Extended Data. Fig. 8a–f). We examined the voltage transfer characteristics (VTC) for the supply voltage ($V_{dd}$). The mid-gate and top-gate of the vertical CMOS were connected to form the input terminal, and the input voltage ($V_{in}$) was varied from 0 to 1.5 V. The output voltage ($V_{out}$) was measured at the output terminal formed by connecting the drain electrodes of nMOS and pMOS transistors. To quantitatively assess the performance of the vertical inverter circuits, we estimated average voltage gain ($A_{gain}$) and noise margin ($NM = NM_L + NM_H$) values from the VTC curves (inset of Fig. 4a shows the voltage gain at $V_{dd}$ = 1 V). As shown in the voltage gain and noise margin as a function of $V_{dd}$, the average $A_{gain}$ and $NM$ values of our vertical inverters are superior to the values reported for TMD-based inverters through stacking owing to seamless stacking during growth (Fig. 4b).

Furthermore, a pull-up and pull-down network were established using two pMOS transistors and two nMOS transistors, respectively, to create NAND and NOR gates (see Extended Data Fig. 8g,h). As depicted in Fig. 4c, successful NAND (red line) and NOR (blue line) functionalities have been achieved out of our vertical inverters. Using data obtained from our vertical CMOS, we conducted a 3D finite element method (FEM) simulation to assess the suitability of our vertical CMOS for advanced node CMOS (Fig. 4d). Throughout the simulation, we predicted the on-current values while scaling down the vertical CMOS (Supplementary Table 2, Table 3, and Table 4 for details). We systematically adjusted the channel length ($L_{ch}$), effective oxide thickness (EOT), and supply voltage ($V_{dd}$) in accordance with the scaling guidelines aligned with the silicon CMOS process node's roadmap. Detailed simulation outcomes can be found in (Extended Data Fig. 9). Due to an unoptimized interface between semiconductors and gate oxide, the estimated on-current value at the 3 nm node is considerably lower than that required by an IRDS roadmap. Here, the estimated density of interface trap ($D_{it}$) of our devices ranges on the order of $10^{13}$ cm$^{-2}$ (pMOS: $8.1\times10^{13}$ cm$^{-2}$, nMOS: $7.0\times10^{13}$ cm$^{-2}$; see Extended Data Fig. 10). However, the calculation predicts that by improving the interface quality with a $D_{it}$ value up to $10^{12}$ cm$^{-2}$, the on-current can substantially exceed the value required by the IRDS (Fig. 4e).[26,27] This implies an evident future direction for research on vertical CMOS.

In conclusion, we have successfully demonstrated a method for arranging single-crystalline semiconductors between amorphous or polycrystalline interlayers through growth at temperatures below 400 ºC, marking the first instance of such an approach. This breakthrough enables the seamless monolithic integration of nMOS and pMOS vertically, resulting in operational vertical inverters. This technology holds the potential to significantly reduce interconnection distances, thereby mitigating RC delays and doubling transistor density within a given wafer space. We believe that the discovered features of this seamless M3D approach can be similarly leveraged for the efficient construction of 3D structures for modern electronic and optoelectronic components.

## Methods

**DFT calculation.** To calculate binding energy of a nucleus under various conditions, diverse slab models incorporating different substrates and flake alignments (vertical or basal) were employed. The vertical distance of a $WSe_2$ from the substrate was set at a value optimizing the binding energy of the largest simulated $WSe_2$ for each substrate (2.0 Å with a-$SiO_2$, 1.0 Å with a-$HfO_2$, and c-$HfO_2$). To prevent asymptotic interactions with atoms possessing dangling bonds, the substrate surface was H-passivated. DFT calculations were performed using the Vienna Ab-initio Simulation Package (VASP) to identify a local minimum state through self-consistent minimization[28,29]. Convergence to the local minimum state was achieved with a threshold value of $10^{-5}$ eV in energy change. Additionally, a dispersion correction (VASP INCAR tag: IVDW = 11) was applied to estimate vdW binding energy[30].

**Synthesis of $WSe_2$ and $MoS_2$.** Confined TMDs were synthesized in a two zone CVD system with a 4-inch quartz tube. In (zone I), 600 mg of Se (or S) powders, and in (zone II), 35 mg of $WO_3$ (or $MoO_3$) powders were placed, with a fixed distance of 33 cm between them. To enhance the uniformity of confined TMD growth at low temperatures, a Quartz-type showerhead with 5 mm holes and spacing was positioned 2 cm above the $WO_3$ (or $MoO_3$) powders. On top of the showerhead, $SiO_2$ trench samples were placed 2 mm above. Confined TMDs were synthesized under atmospheric pressure using Ar (50 sccm)/$H_2$ (50 sccm) as the carrier gas. In (zone I), Se (or S) was maintained at 345 (or 200 °C) for 14 (or 25 min), and in (zone II), $WO_3$ (or $MoO_3$) was maintained at 385 °C for 14 (or 25 min). To prevent cross-contamination, $WSe_2$ and $MoS_2$ were synthesized in separate CVD system, and detailed information on confined growth is described in our previous research[13].

**Characterization of TMDs and devices.** Raman and PL spectra were acquired utilizing a Renishaw InVia Reflex micro-spectrometer equipped with a 532-nm laser. A holographic grating with 2,400 grooves per millimeter dispersed the light. SEM images were captured using a high-resolution SEM (ZEISS Merlin) with an in-Lens detector. The imaging parameters included a 5 mm working distance, an accelerating voltage of 2.5 kV, and a probe current of 85 pA. The structural and elemental analyses of the vertical CMOS were conducted using HRTEM (JEM ARM 200F) and EDS mapping (GIF Quantum ER system) with an accelerating voltage of 200 kV.

**Fabrication and measurements of device and circuit array.** For fabricating vertical CMOS array, at first, a 10 nm-thick a-$HfO_2$ layer was deposited on $SiO_2$/Si wafer using an atomic layer deposition (ALD) process. On the a-$HfO_2$ layer, polymethyl methacrylate (PMMA) A4 950K was spin-coated at 3500 rpm for 60 s and baked at 180°C for 120 s, and PMMA A4 495K was spin-coated and baked using the same method. E-beam lithography (EBL) process without aligning step was followed for patterning confined pocket trench array. After developing the bi-layered PMMA, a 15 nm-thick $SiO_2$ were deposited using an E-beam evaporator, and the outside of the confined pocket trench array regions was removed by a lift-off process. Single-crystalline $WSe_2$ channel layer was then synthesized at 485 $^{o}C$ using the aforementioend synthesis method. The align marks for align during EBL process were patterned using EBL process, and then 10 nm-thick Ti and 150 nm-thick Ni were deposited using E-beam evaporation, followed by lift-off process. Next, drain and source contact regions with a length of 400 nm and a width of 700 nm were patterned using an EBL process with aligning step, followed by depositing 15 nm-thick Pt layers using an E-beam evaporator. The outside of the source/drain contact metal regions was removed by a lift-off process. On top of the single-crystalline $WSe_2$ channel and source/drain electrodes, a 10 nm-thick a-$HfO_2$ was deposited as gate dielectric layer using the ALD process. Using the same EBL, E-beam evaporation, and lift-off processes, a 15 nm-thick Pt gate metal regions were defined. A 10 nm-thick a-$HfO_2$ layers as gate dielectric layers were then deposited using the ALD process, implementing lower pMOS array layer. Next, on top of the pMOS array layer, confined pocket trench array were formed using the EBL, E-beam evaporation, and lift-off processes, followed by synthesizing single-crystalline $MoS_2$ channel layer at 385 $^{o}C$. Drain and source contact regions with a length of 400 nm and a width of 700 nm were patterned using an EBL process, followed by defining 15 nm-thick Cr layers using an E-beam evaporation and lift-off processes. Then, a 20 nm-thick a-$HfO_2$ layer as gate dielectric layer was deposited using the ALD process. Finally, using the EBL, E-Beam evaporation, and lift-off processes, 15 nm-thick Pt gate metal regions were defined, implementing single-crystalline vertical CMOS arrays. The current–voltage characteristics were measured with an B1500A. All the measurements were conducted at room temperature in the air.

## Extended Data

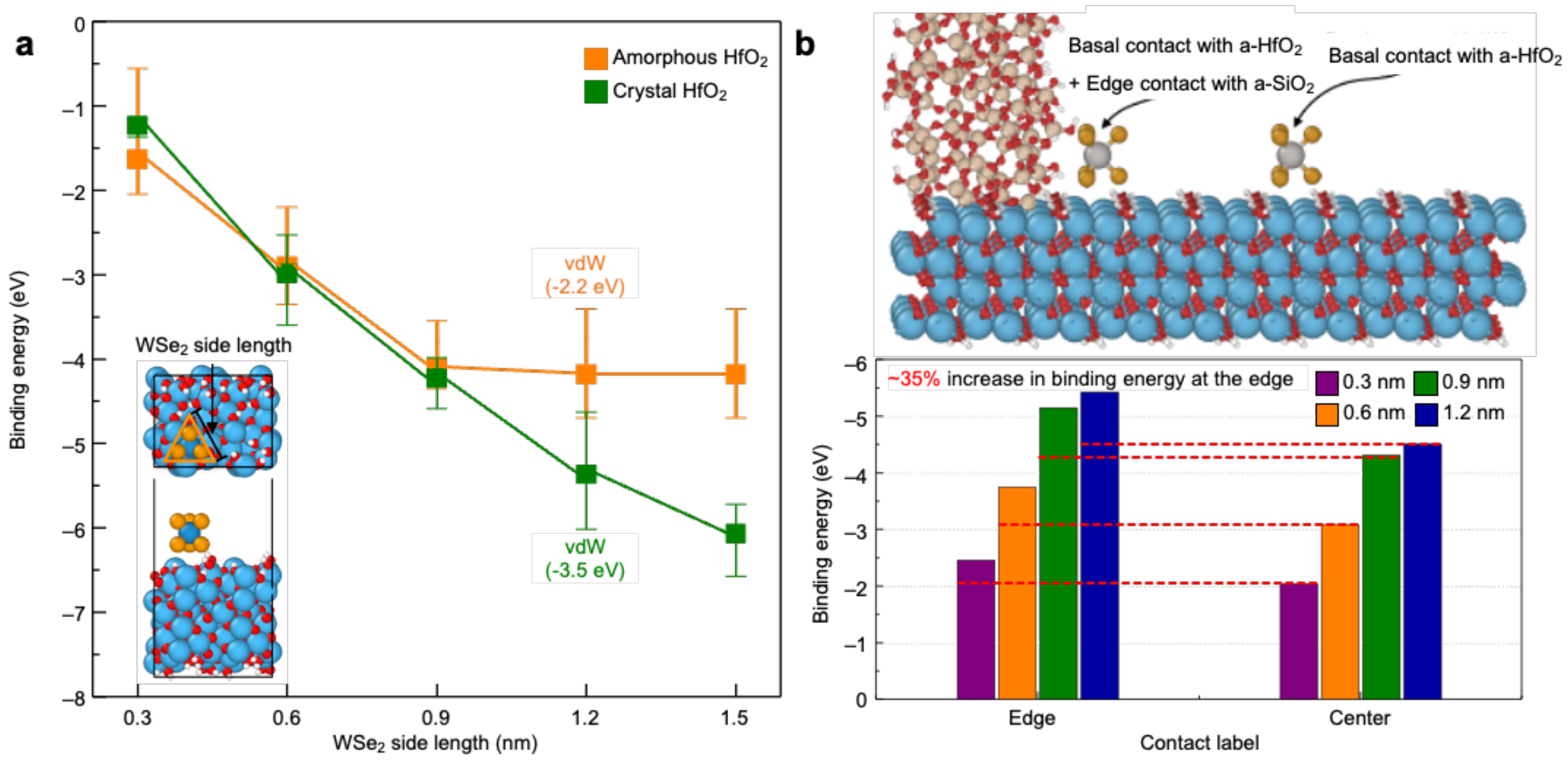

**Extended Data Fig. 1 | Binding character of $WSe_2$ on $HfO_2$. a,** Calculated binding energy of between $WSe_2$ and a-$HfO_2$ (orange-colored plots) or c-$HfO_2$ (green-colored plots) with respect to $WSe_2$ side length. **b,** Schematic showing two nucleation scenario: i) Basal contact with a-$HfO_2$ + Edge contact with $SiO_2$ and ii) Only basal contact with a-$HfO_2$ (top panel), and calculated binding energy with respect to $WSe_2$ side length under edge and center contacted condition (top panel).

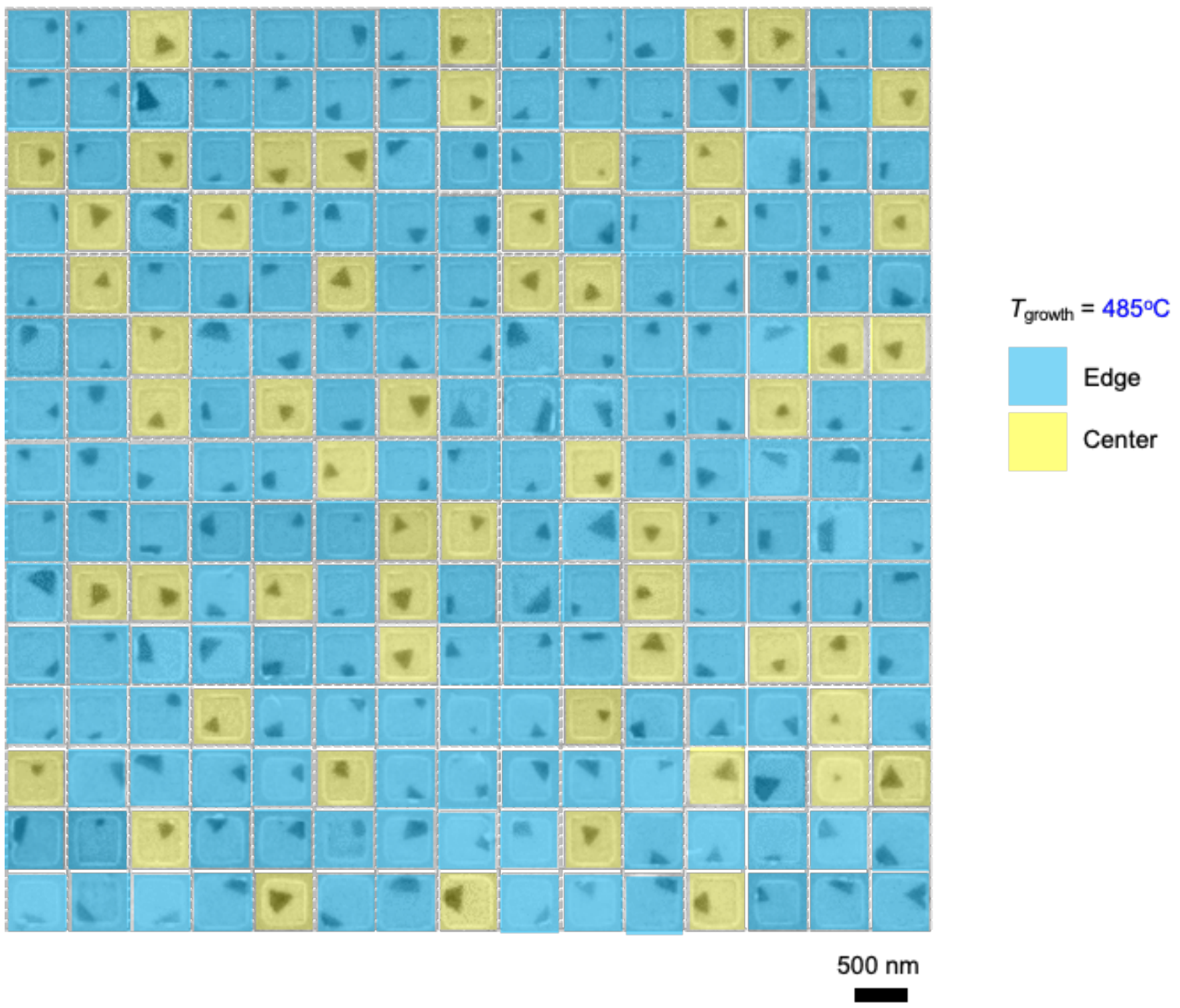

**Extended Data Fig. 2 | Statistics of single-crystalline $WSe_2$ grown in 500 nm-size trench patterns at 485 °C.** Each denoted region indicates initial nuclei at edge (blue-color) and center (yellow-color) of the trench, respectively.

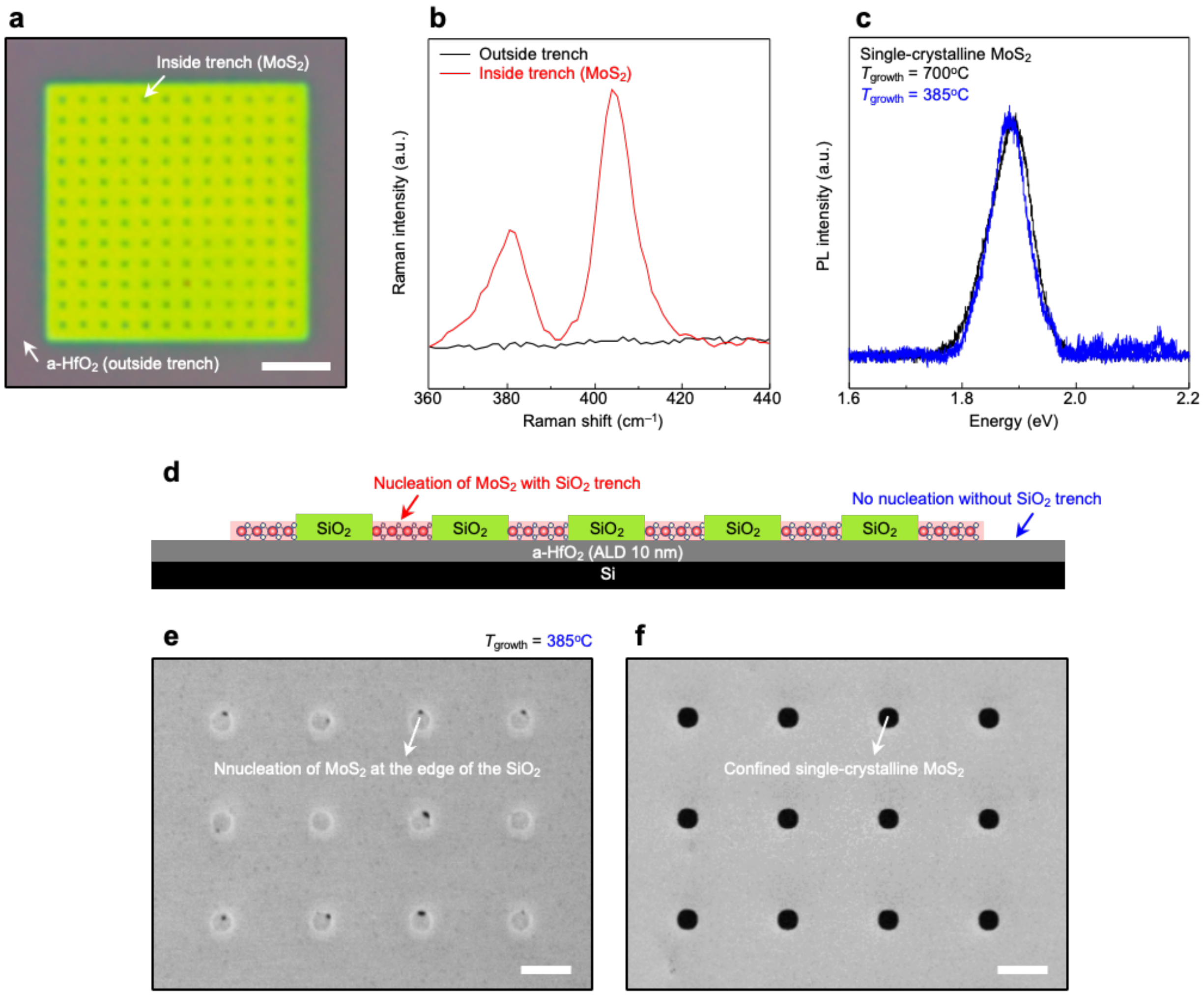

**Extended Data Fig. 3 | Low-temperature growth of single-crystalline $MoS_2$. a,** OM image showing formed $SiO_2$ trench on $HfO_2$ substrate, where scale bar denotes 5 µm. **b,** Raman spectra investigated outside (black-colored profile) and inside (red-colored profile) of trench, respectively, showing nucleation and formation of $MoS_2$ occurring only within trench interiors. **c,** PL spectra investigated on single-crystalline $MoS_2$ grown at 700 °C (black-colored profile) and 385 °C (blue-colored profile), respectively. **d,** Schematic diagram illustrating that nucleation predominantly initiates at edge of $SiO_2$ trenches on $HfO_2$ substrate, rather than in regions without $SiO_2$ trenches. **e–f,** SEM images verifying early-stage nucleation of single-crystalline $MoS_2$ (**e**) and completed growth where single-crystalline $MoS_2$ fills pockets (**f**), respectively. Here, $T_{growth}$ and scale bars are 385 °C and 2 µm.

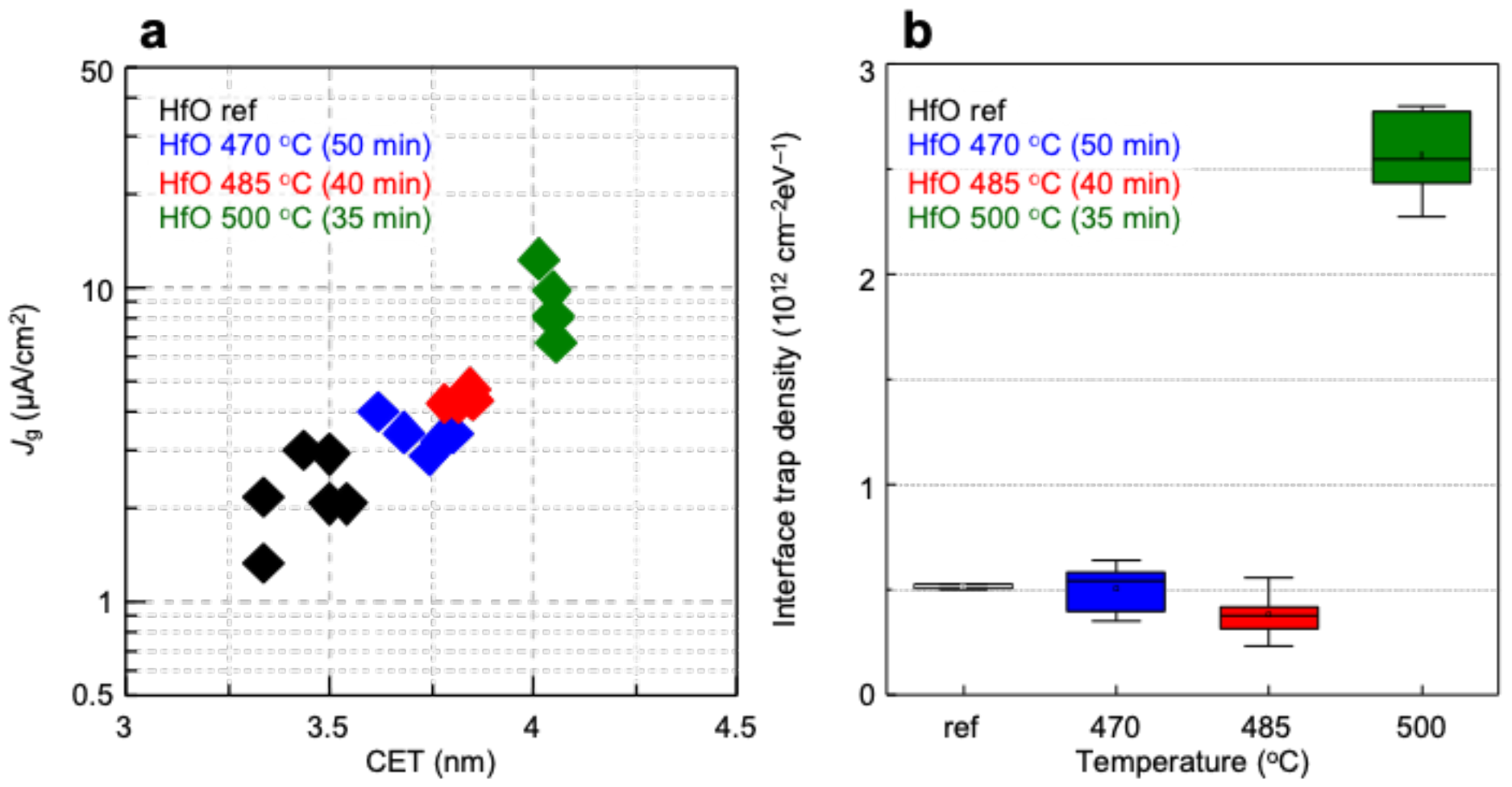

**Extended Data Fig. 4 | Characterization of thermal-treated HfO. a,** Leakage current density of MIM capacitor with respect to temperature of thermal treatment ranging from 470 °C (blue-colored plots) to 485 °C (red-colored plots), and then, to 500 °C (green-colored plots), where insulator is 10 nm HfO. **b,** Interface trap density with respect to temperature of thermal treatment.

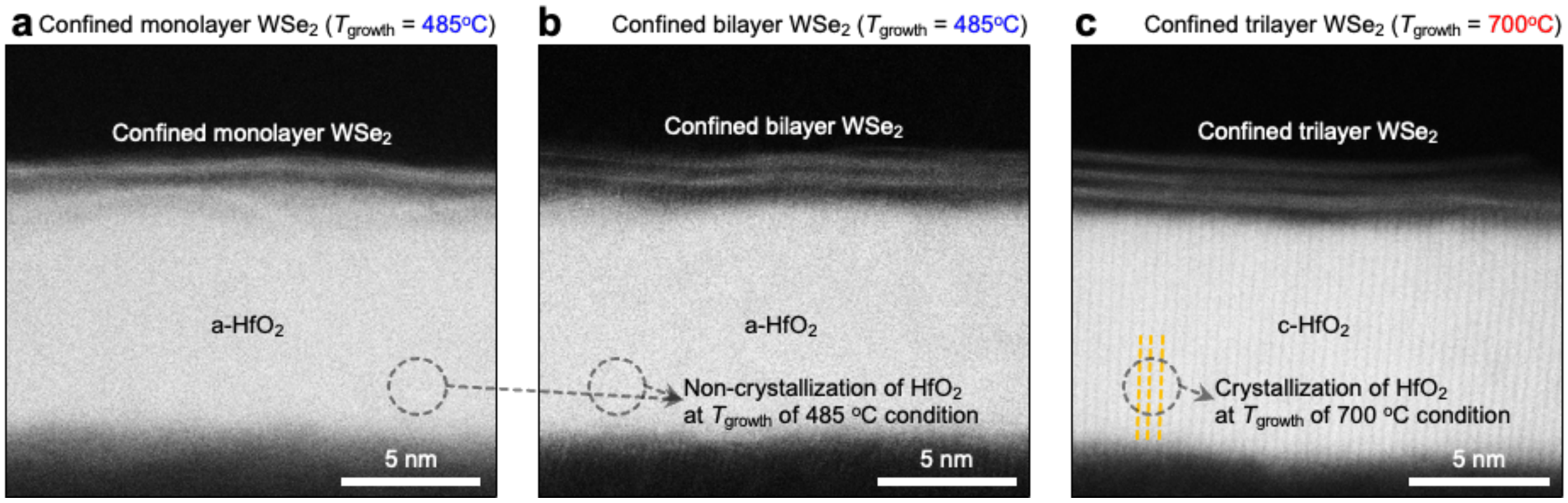

**Extended Data Fig. 5 | TEM analysis on confined $WSe_2/HfO_2$ structure. a–c,** TEM images showing confined $WSe_2/HfO_2$ structure, where monolayer, bilayer, and trilayer $WSe_2$ are grown at $T_{growth}$ of 485 ºC (**a** and **b**; for monolayer and bilayer $WSe_2$, respectively) and 700 ºC (**c**; for trilayer $WSe_2$), respectively. Particularly noteworthy is non-crystallization and crystallization of a-$HfO_2$ layer under $T_{growth}$ of 485 and 700 ºC conditions, respectively.

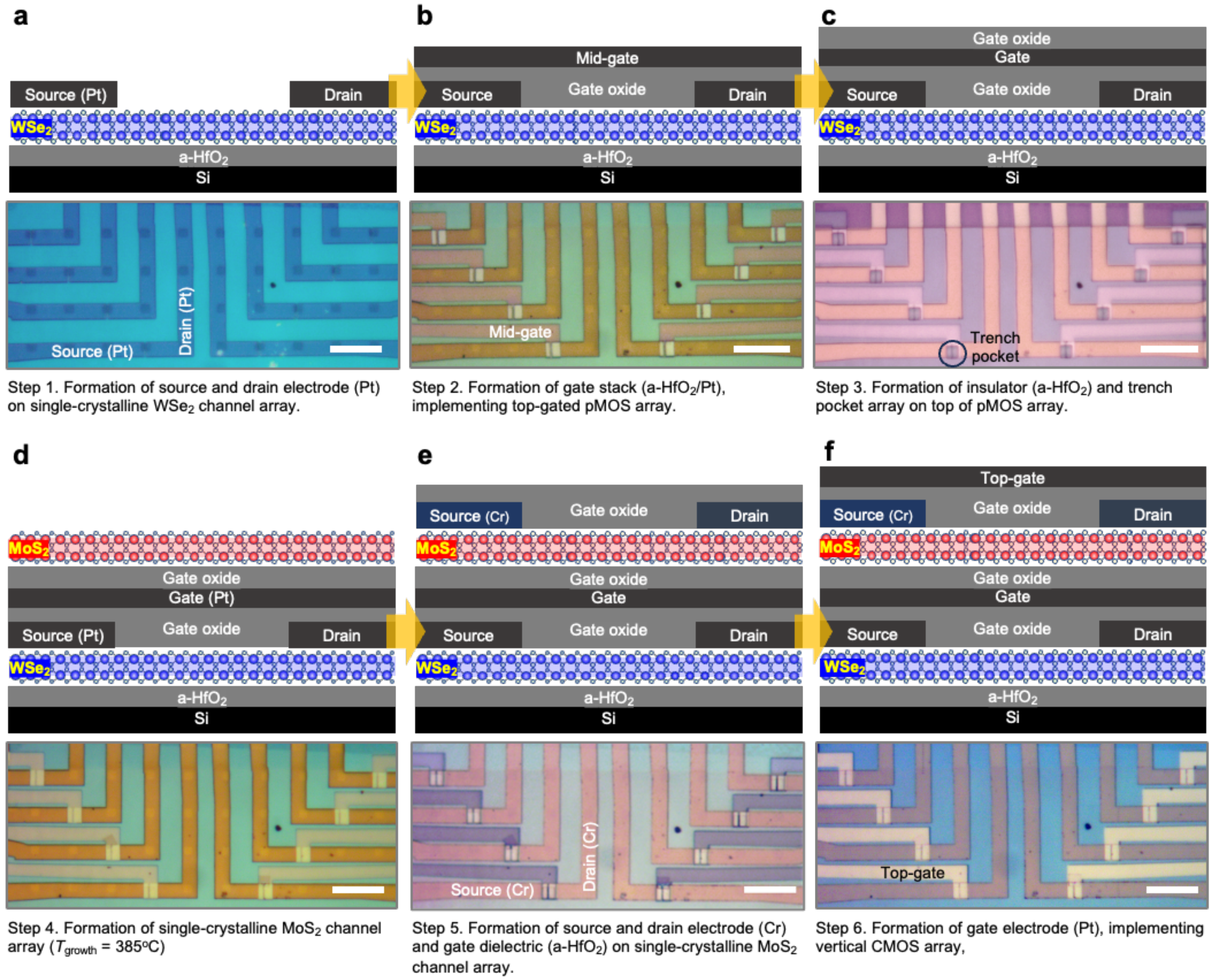

**Extended Data Fig. 6 | Fabrication process of vertical CMOS array.** Schematics (top panel) and OM images (bottom panel) showing fabrication process of vertical CMOS array, where inset scale bars denote 10 µm. Detailed description regarding fabrication process of vertical CMOS array is provided in Method section.

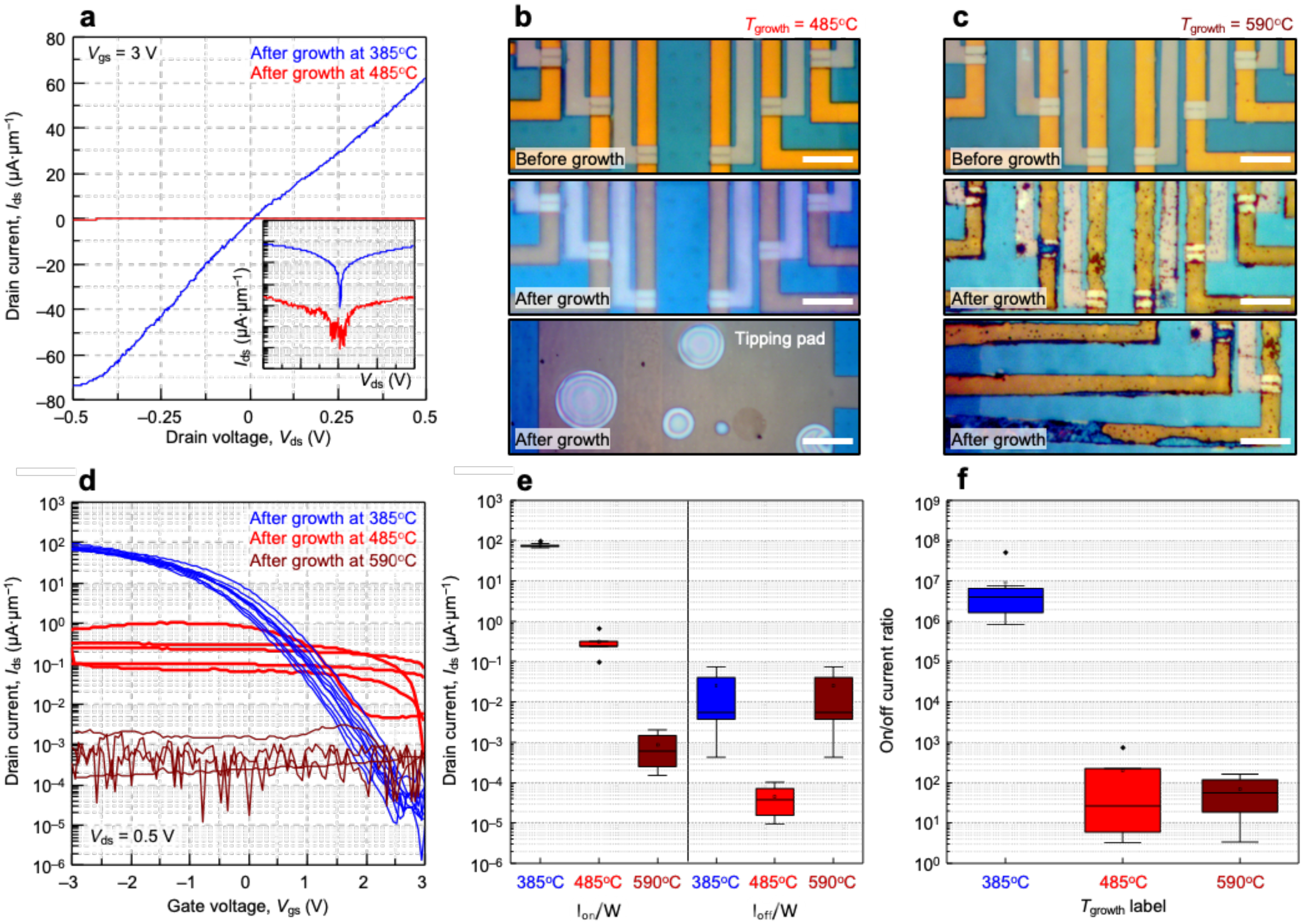

**Extended Data Fig. 7 | Characterization of lower pMOS of vercial CMOS with respect to growth temperature of upper $MoS_2$ channel layer. a–b,** OM images before (top image) and after (middle and bottom images) growth upper $MoS_2$ channel layer at 485 °C (**a**) and 590 °C (**b**), respectively. Scale bars denote 9 μm. **c,** Output characteristics of pMOS measured after growth of $MoS_2$ channel layer at 385 °C (blue-colored line) and 485 °C (red-colored line), respectively. **d,** Transfer characteristics of pMOS measured after growth of $MoS_2$ channel layer at 385 °C (blue-colored line), 485 °C (red-colored line), and 590 °C (wine-colored line), respectively. **e–f,** Statistics of $I_{on}/W$, $I_{on}/W$ (**e**), and on/off current ratio (**f**) with respect to growth temperature ranging from 385 to 485 °C, and then, to 590 °C, which are extracted from transfer characteristic curves. Here, $I_{on}$ and $W$ denote on-current and channel width, respectively, and blanked and blacked dots denote mean and outlier values, respectively.

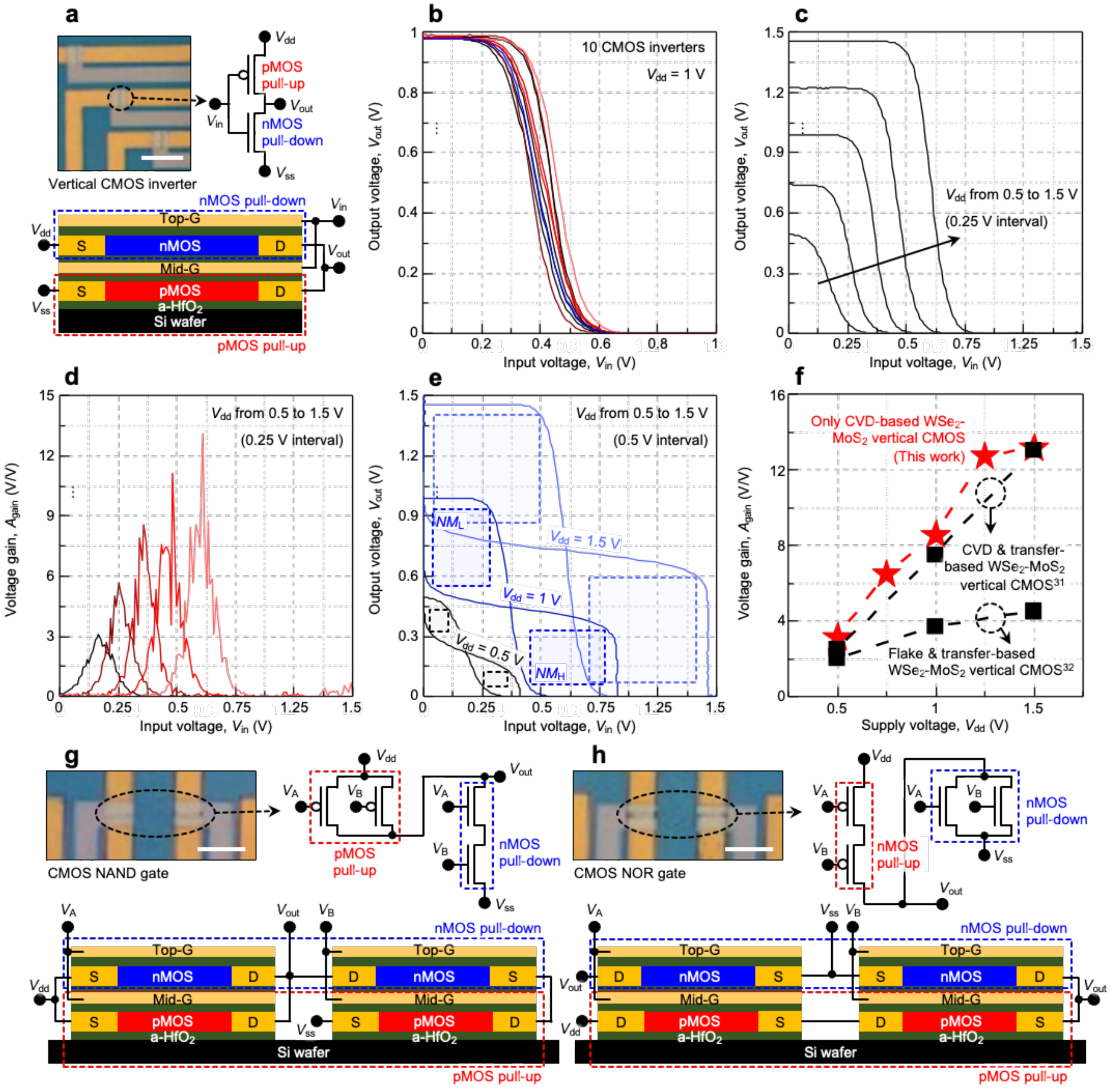

**Extended Data Fig. 8 | Characterization of vertical CMOS-based logic circuits. a,** OM image of fabricated vertical CMOS inverter array (left top panel) and circuit diagram (right top panel) and structural schematic (bottom panel) of vertical CMOS inverter. **b,** Multiple voltage transfer characteristics verified from 10 vertical CMOS inverters. **c,** Voltage transfer characteristics verified under various $V_{dd}$ conditions, ranging from 0.5 to 0.75, 1, 1.25, and 1.5 V. **d,** Extracted voltage gain from voltage transfer characteristics. **e,** Voltage transfer characteristics denoted with noise margin ($NM_L$ and $NM_H$). **f,** Comparison of voltage gain between our vertical CMOS inverter and $WSe_2$-$MoS_2$-based vercial CMOS inverter reported thus far[31,32]. **g–h,** OM image of fabricated vertical NAND (**g**) and NOR (**h**) gates (left top panel) and circuit diagram (right top panel) and structural schematic (bottom panel) of vertical NAND and NOR gates.

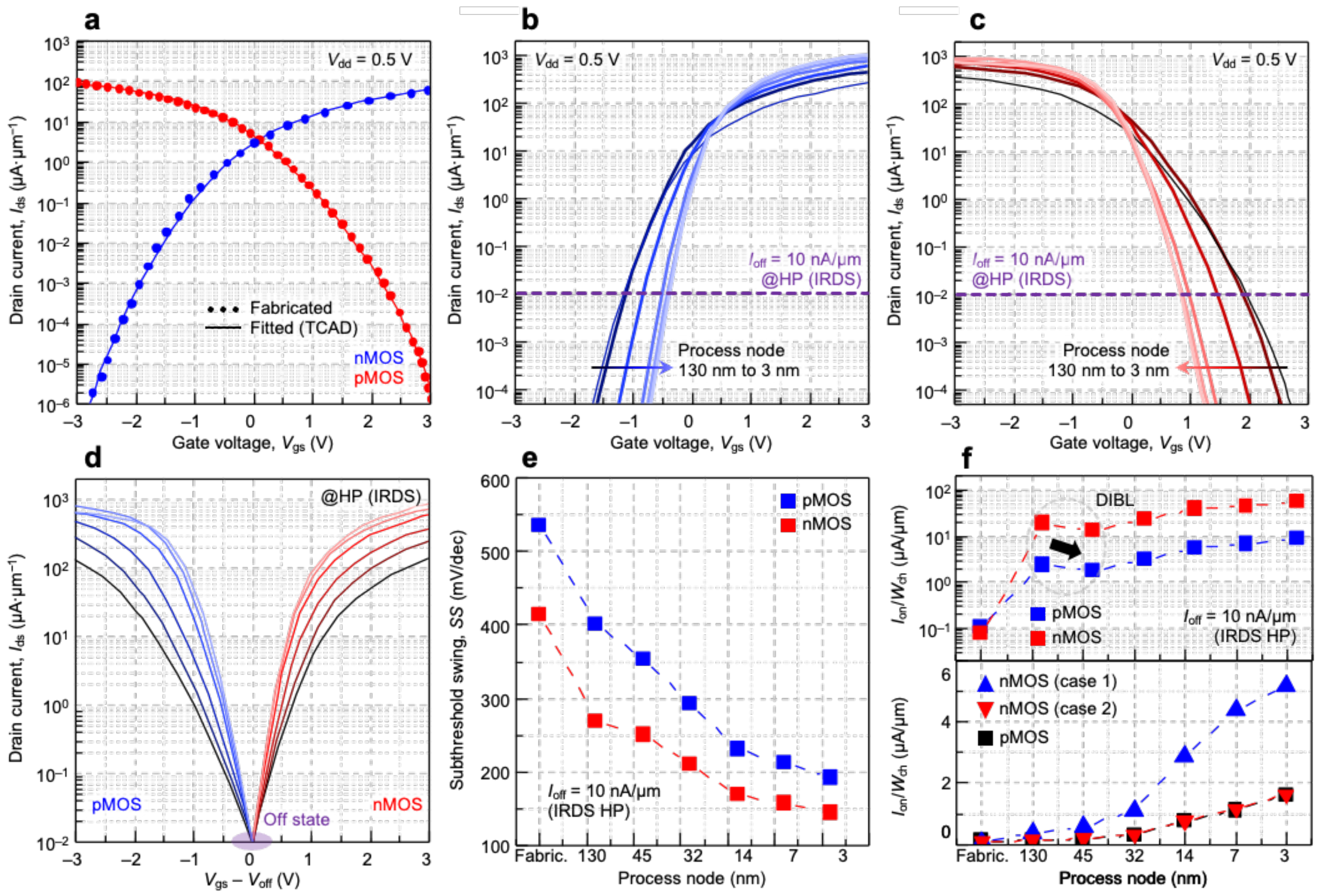

**Extended Data Fig. 9 | Advanced CMOS process node-based vertical CMOS. a,** Transfer characteristics of nMOS and pMOS transistors of vertical CMOS, where dotted and line profiles denote transfer characteristic of experimentally obtained and TCAD-based fitted results, respectively. **b–c,** Predicted transfer characteristics of nMOS (**b**) and pMOS (**c**) transistors of vertical CMOS with respect to CMOS process node ranging from 130 to 45, 32, 14, 7, and 3 nm. **d,** Transfer characteristic of nMOS nad pMOS transistors of vertical CMOS with respect to CMOS process node with fixed with $I_{\mathrm{off}}$ of 10 nA/μm. **e,** Predicted $I_{\mathrm{on}}/W_{\mathrm{ch}}$ values of vertical CMOS with respect to CMOS process node ranging from 130 nm to 45 nm, 32 nm, 14 nm, 7 nm, and 3 nm, respectively, where fixed off current is set to 10 nA/μm. **f,** Predicted $I_{\mathrm{on}}/W_{\mathrm{ch}}$ value of vertical CMOS with respect to CMOS process node ranging from 130 to 45, 32, 14, 7, and 3 nm, respectively, where fixed off current is set to 10 nA/μm (top panel), and predicted $I_{\mathrm{on}}/W_{\mathrm{ch}}$ values of pMOS and nMOS transistors with respect to CMOS process node (bottom panel), without (blue-colored plots) and with (red-colored plots) optimization of $D_{\mathrm{np}}$.

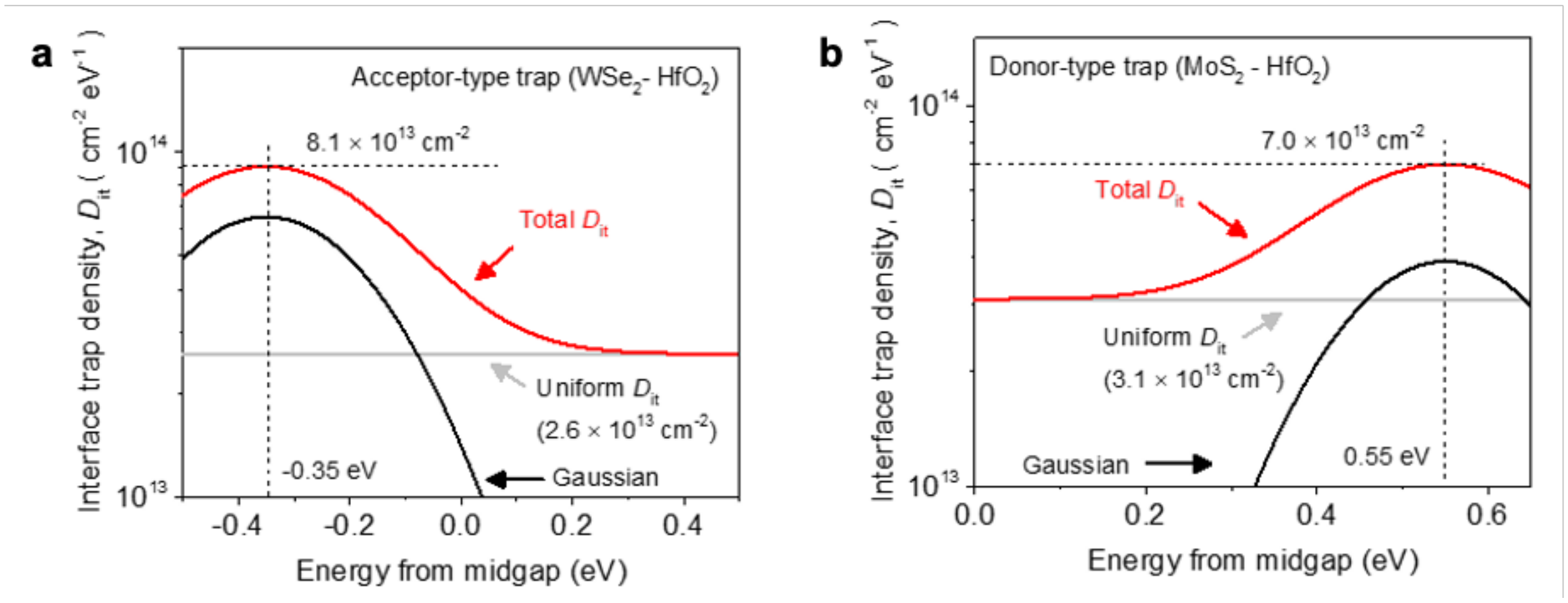

**Extended Data Fig. 10 | $D_{it}$ distribution model of oxide-channel interface extracted through TCAD simulation. a,** $WSe_2$-$HfO_2$. **b,** $MoS_2$-$HfO_2$.

# Supplementary Information

| Materials | Crystallinity | Growth | | | Device | | | | | | Ref |
|---|---|---|---|---|---|---|---|---|---|---|---|
| | | Type | Temperature | Substrate | Gate stack | Contact | Channel dimension | Operational voltage | $I_{on}/W_{ch}$ | Current on/off ratio | |
| $WSe_2$ | Single | CVD | 385 °C | a-$HfO_2$ | 10 nm $HfO_2$ | Pt | 400 nm ($L_{ch}$)/ 700 nm ($W_{ch}$) | 0.5 V | 75.3 μA/μm | ~$10^6$ | This work |
| $MoS_2$ | Single | CVD | 385 °C | a-$HfO_2$ | 10 nm $HfO_2$ (bottom)/ 20 nm $HfO_2$ (top) | Cr | 400 nm ($L_{ch}$)/ 700 nm ($W_{ch}$) | 0.5 V | 67.2 μA/μm | ~$10^7$ | |
| $WSe_2$ | Single | CVD | 890 °C | Sapphire | 10 nm $HfO_2$/ 100 nm $SiO_2$ | Pd | 700 nm ($L_{ch}$)/ 2 μm ($W_{ch}$) | 1 V | 89.9 μA/μm | ~$10^8$ | 1 |
| $WSe_2$ | Poly | CVD | 1180 °C | $SiN_x/SiO_2$ | $HfO_2$ | Sb | 100 nm ($L_{ch}$) | 0.5 V | ~ 35 μA/μm | Not verified (-) | 2 |
| $WSe_2$ | Poly | CVD | 900 °C | $SiO_2$ | 100 nm $SiN_x$ | Bi | 1 μm ($L_{ch}$) | 1 V | 14 μA/μm | ~$10^8$ | 3 |
| $WSe_2$ | Poly | MOCVD | 537 °C | Sapphire | - | - | - | - | - | - | 4 |
| $WSe_2$ | Single | MBE | 200~400 °C | Au/Sapphire | - | - | - | - | - | - | 5 |
| $MoS_2$ | Single | CVD | 750 °C | a-$HfO_2$ | 10 nm $HfO_2$/ 100 nm $SiO_2$ | Ni | 700 nm ($L_{ch}$)/ 2 μm ($W_{ch}$) | 1 V | 86.7 μA/μm | ~$10^7$ | 1 |
| $MoS_2$ | Poly | CVD | 1000 °C | Sapphire | $HfO_2$ | Sb(110) | 200 nm ($L_{ch}$) | 1 V | 1440 μA/μm | ~$10^8$ | 2 |
| $MoS_2$ | Poly | CVD | 625 °C | $SiO_2$ | 100 nm $SiN_x$ | Bi | 35 nm ($L_{ch}$) | 1.5 V | 1135 μA/μm | $10^6$ | 3 |
| $MoS_2$ | Poly | MOCVD | 320 °C | $SiO_2$ | 100 nm $SiN_x$ | Bi | 120 nm ($L_{ch}$) | 1.5 V | 560 μA/μm | $10^7$ | 3 |
| $MoS_2$ | Single | MOCVD | 320 °C | $SiO_2$, Sapphire | 300 nm $SiO_2$ | - | - | 1 V | ~200 μA/μm | ~$10^5$ | 6 |
| $MoS_2$ | Poly | MOCVD | 580 °C | GaN/Si | 30 nm $Al_2O_3$ | Cr | 45 nm ($L_{ch}$)/ 10 μm ($W_{ch}$) | 1 V | 2.2 μA/μm | $10^9$ | 7 |
| $MoS_2$ | Poly | MOCVD | 275 °C | $SiO_2$ | 20 nm $Al_2O_3$ | Au | 200 nm ($L_{ch}$) | 0.7 V | 40 μA/μm | $10^5$ | 8 |

**Supplementary Table 1 | Comparison table of synthesized TMD-based transistors in terms of growth temperature of TMD and performance of device.**[1–8]

| Type | Material | Parameter | Value |
|---|---|---|---|
| Channel | $WSe_2$ | Dielectric constant of in-plane ($\varepsilon_{\parallel}$) | 15.7 |
| | | Dielectric constant of out-of-plane ($\varepsilon_{\perp}$) | 7.6 |
| | | Energy band gap ($E_g$) | 1.15 [eV] |
| | | Electron affinity ($\chi$) | 4.15 [eV] |
| | | Intrinsic Fermi level ($E_{Fi}$) | 4.78 [eV] |
| | | Effective mass of electron ($m_e^*/m_0$) | 0.33 |
| | | Effective mass of hole ($m_h^*/m_0$) | 0.46 |
| | $MoS_2$ | Dielectric constant of in-plane ($\varepsilon_{\parallel}$) | 15.7 |
| | | Dielectric constant of out-of-plane ($\varepsilon_{\perp}$) | 6.6 |
| | | Energy band gap ($E$) | 1.27 [eV] |
| | | Electron affinity ($\chi$) | 4.52 [eV] |
| | | Intrinsic Fermi level ($E_{Fi}$) | 5.16 [eV] |
| | | Effective mass of electron ($m_e^*/m_0$) | 0.56 |
| | | Effective mass of hole ($m_h^*/m_0$) | 0.62 |
| Metal | Pt | Fermi level ($E_F$) | 5.15 [eV] |
| | Cr | Fermi level ($E_F$) | 4.66 [eV] |
| | Pd | Fermi level ($E_F$) | 5.22 [eV] |

**Supplementary Table 2 | Electrical parameter for TCAD simulation.**[9–12]

| | Fabric. | 130 nm | 45 nm | 32 nm | 14 nm | 7 nm | 3 nm |
|---|---|---|---|---|---|---|---|
| EOT (nm) | 1.56 | 1.45 | 1.26 | 1.0 | 0.73 | 0.64 | 0.54 |
| Lg (nm) | 400 | 65 | 27 | 24 | 24 | 20 | 16 |
| $V_{dd}$ (V) | 0.5 | 1.2 | 0.97 | 0.9 | 0.8 | 0.75 | 0.7 |

**Supplementary Table 3 | Ground rules roadmap for logic devices.[13,14]**

| **Case 1** | **Fabric.** | **130 nm** | **45 nm** | **32 nm** | **14 nm** | **7 nm** | **3 nm** |
|---|---|---|---|---|---|---|---|
| Tox_p | 20 | 9.3 | 8.1 | 6.4 | 4.7 | 4.1 | 3.5 |
| Tox_n | 10 | 9.3 | 8.1 | 6.4 | 4.7 | 4.1 | 3.5 |
| Dnp | 10 | 9.3 | 8.1 | 6.4 | 4.7 | 4.1 | 3.5 |
| **Case 2** | **Fabric.** | **130 nm** | **45 nm** | **32 nm** | **14 nm** | **7 nm** | **3 nm** |
| Tox_p | 20 | 9.3 | 9.1 | 6.4 | 4.7 | 4.1 | 3.5 |
| $T_{ox_n}$ | 10 | 9.3 | 9.1 | 6.4 | 4.7 | 4.1 | 3.5 |
| $D_{np}$ | 10 | 28.8 | 21.2 | 16.0 | 12.2 | 9.6 | 8.0 |

**Supplementary Table 4 | Geometric parameter values for Case 1 and Case 2.**

**Supplementary References**